\documentclass[aps,pra,english,twocolumn,letterpaper,notitlepage]{revtex4-1}

\usepackage[T1]{fontenc}
\usepackage{graphicx}
\usepackage{epstopdf}
\usepackage{amssymb}
\usepackage{amsmath}
\usepackage[urlcolor=blue,hyperindex,colorlinks,bookmarks=true,linkcolor=red,citecolor=black,]{hyperref}
\usepackage[normalem]{ulem}
\usepackage[abs]{overpic}
\usepackage{color}
\usepackage{rotating}
\usepackage{subfigure}
\usepackage{physics}
\usepackage{bm}
\usepackage{enumerate}
\usepackage[utf8]{inputenc}
\usepackage{booktabs}
\usepackage{placeins}

\usepackage[english]{babel}
\usepackage{blindtext,tikz}
\usetikzlibrary{calc}

\def\iden{\mathbb{I}}
\def\idenp{\mathcal{I}}

\newcommand{\qp}{\mathcal{E}}

\begin{document}

\title{Bootstrapping quantum process tomography via a perturbative ansatz}

\author{L.~C.~G.~Govia}
\email{luke.c.govia@raytheon.com}
\author{G.~J.~Ribeill}
\author{D.~Rist\`e}
\author{M.~Ware}
\author{H.~Krovi}
\affiliation{Raytheon BBN Technologies, 10 Moulton St., Cambridge, MA 02138, USA}


\begin{abstract}
  Quantum process tomography has become increasingly critical as the need grows for robust verification and validation of candidate quantum processors. Here, we present an approach for efficient quantum process tomography that uses a physically motivated ansatz for an unknown quantum process. Our ansatz bootstraps to an effective description for an unknown process on a multi-qubit processor from pairwise two-qubit tomographic data. Further, our approach can inherit insensitivity to system preparation and measurement error from the two-qubit tomography scheme. We benchmark our approach using numerical simulation of noisy three-qubit gates, and show that it produces highly accurate characterizations of quantum processes.  Further, we demonstrate our approach experimentally, building three-qubit gate reconstructions from two-qubit tomographic data.
\end{abstract}

\maketitle

\section{Introduction}

Recent years have seen remarkable progress in quantum information processing, with rapid advancement towards high-fidelity multi-qubit systems \cite{Ballance:2016aa,Barends:2014aa,Zajac439}, some of which are now publicly available \cite{IBMQ,Rigetti}. This has enabled significant achievements in many aspects of quantum computation, such as first demonstrations of the building blocks for error correction and fault-tolerance, e.g.~\cite{Nigg302,Kelly:2015aa,Ofek:2016aa,Takita:2017aa,Linkee1701074,Rosenblum266,Hu:2019aa}. Concurrently, demonstrations of noisy-intermediate-scale quantum algorithms \cite{Preskill:2018aa} that do not require full fault-tolerance, e.g.~\cite{Peruzzo:2014aa,OMalley:2016aa,Riste:2017aa,Kandala:2017aa,Colless:2018aa}, make real world applications of quantum information processing a near-term possibility.

In light of these achievements, the need for robust, accurate, and efficient validation and verification of quantum processors becomes ever more pressing. This is the natural domain of \emph{quantum state tomography} (QST) and \emph{quantum process tomography} (QPT). Respectively, QST and QPT seek to characterize the state of a quantum processor or the dynamical map of its evolution \cite{Nielsen00}. Unfortunately, na\"ive implementations of both QST and QPT require measuring a set of observables, and the size of this set scales exponentially with the number of qubits. For practical purposes, this scaling has limited full QST and QPT to small system sizes, e.g.~\cite{Walther:2005aa,Bialczak:2010aa}, though this can be improved using approximate characterizations \cite{Shabani:2011aa,Lanyon:2017aa}, or in situations with large amounts of symmetry \cite{Medeiros-de-Araujo:2014aa,Yokoyama:2013aa}.

Further compounding QPT, the most error-prone operations are often system preparation and measurement (SPAM), which can overwhelm the intrinsic error in high-fidelity quantum processes and hinder their characterization. Several SPAM-insensitive metrics exist, such as the widely-successful randomized benchmarking \cite{Emerson:2005aa,Knill:2008aa,Magesan:2011aa,Magesan:2012ab} and its variants \cite{Magesan:2012aa,Gambetta:2012aa,Carignan-Dugas:2015aa,Chasseur:2017aa,Wood:2018aa,Helsen:aa,Proctor:aa}, as well as gate set tomography (GST) \cite{Merkel:2013aa,Greenbaum:aa,Blume-Kohout:2017aa}. Randomized benchmarking has the additional benefit of overcoming the exponential scaling of standard QPT, but at the cost of returning only a single number characterizing the quantum process.

In this work, we present an approach to efficient QPT that reduces the exponential scaling to quadratic scaling, while still returning a full process matrix describing the quantum process. We propose the \emph{Pairwise Perturbative Ansatz} (PAPA), which describes the unknown quantum process as sequential two-qubit processes on all qubit pairs. We show how to fit the free parameters of our ansatz to data obtained from QPT of two-qubit subsets of the full system. When this data is provided by SPAM-insensitive tomography, such as GST, our approach becomes SPAM-insensitive as well as efficient.

The paper is organized as follows. In section~\ref{sec:Background} we provide background information on QPT and compare PAPA to existing QPT protocols. In sections~\ref{sec:LA} and~\ref{sec:Char} we describe PAPA in detail, and outline how to obtain the necessary tomographic data to obtain a PAPA characterization. Section~\ref{sec:Exp} details an experimental demonstration of the PAPA process.  In section~\ref{sec:Sim} we benchmark the PAPA approach using numerical simulation, and finally in section~\ref{sec:Conc} we present our conclusions.

\section{Background}
\label{sec:Background}

A generic $N$-qubit quantum process, which we label as $\qp$, has $16^N - 4^N$ free parameters, and the goal of QPT is to determine these free parameters. This makes na\"ive QPT an exponentially hard problem, as an exponential number of measurement settings (unique observables) are required to determine the free parameters. Even for small to modest $N$ this scaling is practically unfavorable, and QPT is very challenging experimentally.

Process tomography can be rephrased as state tomography of the Choi dual-state (via the Choi-Jamiołkowski isomorphism), which is the state formed when the unknown process acts on one half of a maximally entangled state in a Hilbert space of dimension $2^{2N}$, given by
\begin{align}
  \rho_{\qp} &= \frac{1}{2^N}\sum_{\mu\nu}\ketbra{\psi_\mu}{\psi_\nu} \otimes \qp\left(\ketbra{\psi_\mu}{\psi_\nu}\right),
\end{align}
where $\{\ket{\psi_\mu}\}$ is an orthonormal basis for $N$-qubit Hilbert space.

Thus, one can use efficient state tomography methods for process tomography, such as compressed sensing \cite{Gross:2010aa,Gross:2011aa,Shabani:2011aa} and matrix-product-state (MPS) parameterizations \cite{Baumgratz:2013aa,Cramer:2010aa,Holzapfel:2015aa,Lanyon:2017aa}. Unfortunately, the matrix completion algorithms that underly these approaches can themselves be inefficient in run-time. This issue can be circumvented using constrained approaches, as in Refs.~\cite{Cramer:2010aa,Lanyon:2017aa}, which restrict to pure state descriptions of the unknown quantum state.

Both compressed sensing and MPS parameterizations implicitly assume an ansatz for the unknown quantum process, that it is either low rank, or has a matrix product structure (and thus correlations are not long range) respectively. Our pairwise perturbative ansatz assumes a different physical constraint on the unknown process: {\it that it is intrinsically built from two-qubit processes on all pairs of qubits}. Like the MPS approach, this implies that few-body QPT is sufficient to find a PAPA characterization of the unknown process. Unlike an MPS, PAPA has no locality constraint on correlations, and allows for long-range correlations, though these come about only via local interactions between qubit pairs. Further, we will see in the next section that the PAPA constraint is physically motivated, unlike the low rank restriction of compressed sensing.

\section{Ansatz for Process Tomography}
\label{sec:LA}

\begin{figure}
  \includegraphics[width=\columnwidth]{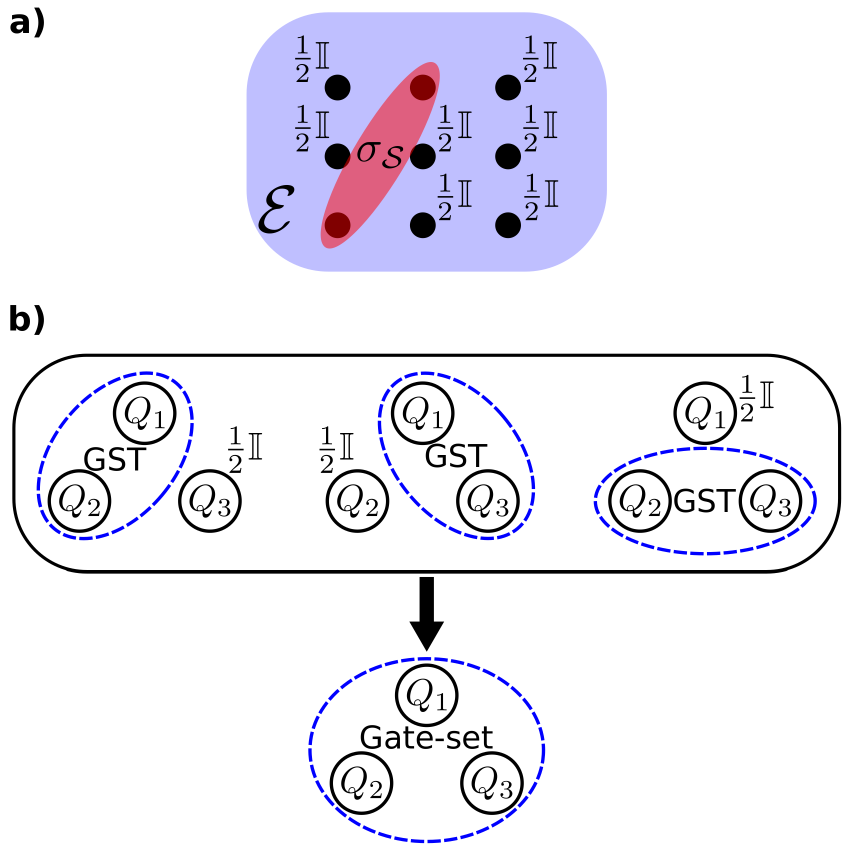}
  \caption{{\bf a)} Pairwise Perturbative Ansatz (PAPA) tomography: for all qubit pairs, characterize the effective two-qubit process (Choi state $\sigma_{\mathcal{S}}$) when the unknown $N$-qubit process $\qp$ occurs, and all other qubits start in the maximally mixed state. {\bf b)} three-qubit PAPA+GST: characterized two-qubit gate sets are bootstrapped to a three-qubit gate set via PAPA.}
  \label{fig:PAPA}
\end{figure}

We propose to restrict the unknown Choi state by assuming an ansatz for its form. This restricts the number of free parameters in the unknown process {\it a priori}, and therefore restricts the number of measurement settings required.

We will assume an ansatz where the unknown $N$-qubit process is written as a composition of two-qubit processes, consisting of quantum processes for each qubit pair in the system. This is most easily expressed in terms of the super-operator matrix representation $\hat{\qp}$ of the quantum process $\qp$, as the series composition becomes a product of matrices. This has the general form
\begin{align}
    \hat{\qp} = \prod_{k=1,n=1}^{N-1,N-k}\hat{\qp}_{k,n+k}, \label{eqn:2QPar}
\end{align}
where $\qp_{k,n+k}$ is an arbitrary two-qubit process on qubits $k$ and $(n+k)$ with no restrictions.

The product runs over all pairs of qubits, of which there are $(N^2-N)/2$. Each of the unknown two-qubit processes can be written as
\begin{align}
  \nonumber\qp_{k,n+k} = \sum^{16}_{i_{k,n},j_{k,n}}\chi_{i_{k,n}}^{j_{k,n}}\Big(&\idenp^{\otimes k-1}\otimes\mathcal{A}^{(k)}_{i_{k,n}}\otimes\idenp^{\otimes n-1}\\
  &\otimes\mathcal{A}^{(k+n)}_{j_{k,n}}\otimes \idenp^{\otimes N-k-n}\Big), \label{eqn:2Qlocal}
\end{align}
where $\{\mathcal{A}^{(k)}_{i_{k,n}}\}$ is a complete basis for single-qubit processes and $\idenp$ is the identity process. $\chi_{i_{k,n}}^{j_{k,n}}$ is an element of the $\chi$-matrix describing the two-qubit process, and the summation variables $i_{k,n}$ and $j_{k,n}$ are subscripted to emphasize that they correspond to a particular qubit pair.

There are many possible ansatz for an unknown quantum process \cite{Gross:2010aa,Gross:2011aa,Shabani:2011aa,Baumgratz:2013aa,Cramer:2010aa,Holzapfel:2015aa,Lanyon:2017aa}, but the form we have chosen is particularly well motivated physically. As it is the composition of two-qubit processes in sequence, it captures the natural two-body quantum operations that occur in a gate-based quantum computation. It can completely specify any \emph{ideal} gate operation (single-layer quantum circuit built from one and two-qubit gates), and will contain both single-qubit errors and correlated two-qubit errors as independent free parameters. It also describes processes that involve more than two-qubits, but as combinations of two-qubit processes performed in sequence. Thus, it describes general processes in a perturbative fashion, built from one- and two-qubit processes.

While each arbitrary two-qubit process described by Eq.~\eqref{eqn:2Qlocal} is parameterized in terms of a basis with $16^2$ elements, its $\chi$-matrix has only $16^2-4^2 = 240$ free parameters. There are ${{N}\choose{2}} = (N^2-N)/2$ two-qubit subsets, and so the total number of free parameters in our ansatz is $120(N^2-N)$. As this scales quadratically with qubit number, PAPA is an efficient approach to QPT.

QPT with PAPA consists of determining the $\chi$-matrix for each two-qubit process in the product in Eq.~\eqref{eqn:2QPar}. Inspired by the local tomography used in \cite{Cramer:2010aa,Lanyon:2017aa}, we will use the tomographic characterization of two-qubit processes on all pairs of qubits to determine these free parameters. In essence, from characterization of two-body processes, we bootstrap to a multi-qubit process of PAPA form.

To compare the PAPA ansatz to two-qubit tomographic data, we must determine a notion of a two-qubit reduction of a process $\qp$. This is most easily done in terms of the Choi state $\rho_\qp$. For the two-qubit subset $\mathcal{S} = \{m, p\}$ this takes the form
\begin{align}
  \rho_{\mathcal{S}}= \frac{1}{2^N}\sum_{\mu\nu}{\rm Tr}_{/\mathcal{S}}\left[\ketbra{\psi_\mu}{\psi_\nu}\right] \otimes {\rm Tr}_{/\mathcal{S}}\left[\qp\left(\ketbra{\psi_\mu}{\psi_\nu}\right)\right],  \label{eqn:RedMat}
\end{align}
where by ${\rm Tr}_{/\mathcal{S}}[\rho]$ we mean the partial trace of all qubits other than those in the set $\mathcal{S}$, and it is important to note that the partial trace is applied to both ``parts'' of the Choi state. Using the orthogonality of the $N$-qubit basis, we see that
\begin{align}
  {\rm Tr}_{/\mathcal{S}}\left[\ketbra{\psi_\mu}{\psi_\nu}\right] = \delta_{\mu_{/\mathcal{S}}, \nu_{/\mathcal{S}}}\ketbra{\psi_{\mu_{\mathcal{S}}}}{\psi_{\nu_{\mathcal{S}}}},
\end{align}
where the indices $\mu_{\mathcal{S}}$ ($\mu_{/\mathcal{S}}$) are the subset of indices in $\mu$ that correspond to the qubits inside (outside) of the subset $\mathcal{S}$. Thus, the reduced Choi state of the unknown process can be written as
\begin{align}
  \nonumber\rho_{\mathcal{S}}= \frac{1}{2^2}\sum_{\mu_{\mathcal{S}}\nu_{\mathcal{S}}}&\Big(\ketbra{\psi_{\mu_{\mathcal{S}}}}{\psi_{\nu_{\mathcal{S}}}} \\ &\otimes {\rm Tr}_{/\mathcal{S}}\left[\qp\left(\ketbra{\psi_{\mu_{\mathcal{S}}}}{\psi_{\nu_{\mathcal{S}}}}\otimes\frac{\mathbb{I}_{N-2}}{2^{N-2}}\right)\right]\Big), \label{eqn:RedMat2}
\end{align}
where $\mathbb{I}_{N-2}$ is the identity matrix of dimension $2^{N-2}$.

To determine the free parameters in the PAPA ansatz, for each pair of qubits we compare the two-qubit reduced Choi states described by Eq.~\eqref{eqn:RedMat2} to the corresponding experimentally characterized two-qubit Choi state. Operationally, this amounts to performing two-qubit QPT on the $(N^2-N)/2$ pairs of qubits. Each of the pairwise characterized two-qubit processes is described by $16^2-4^2 = 240$ complex numbers, which gives a total of $120(N^2-N)$ total complex numbers describing the two-qubit process characterization of all pairs of qubits.

Thus, we have exactly as many constraints (coming from experimental characterization) as there are free parameters in PAPA. This further motivates our choice of ansatz, as we have made use of all available data from two-qubit characterizations of the unknown multi-qubit process. In the following section we complete our description of PAPA tomography by describing what two-qubit processes must be characterized for each qubit pair in order to solve for the unknown parameters in our ansatz.

\section{Characterizing The Two-Qubit Processes}
\label{sec:Char}

In the most general version of QPT, there is a completely unknown quantum process which one wishes to determine. Applying PAPA to this problem, the required two-qubit QPT is derived from the form of Eq.~\eqref{eqn:RedMat2}. For a pair of qubits defined by the subset $\mathcal{S}$ we perform two-qubit QPT to characterize the effective process the qubits in $\mathcal{S}$ experience when the unknown process $\qp$ is implemented on all $N$ qubits (with all other qubits initialized in the maximally mixed state), as depicted in Fig.~\ref{fig:PAPA}a).

To see that Eq.~\eqref{eqn:RedMat2} describes a valid two-qubit process, we describe the unknown $N$-qubit process in a basis of $N$-qubit processes as
\begin{align}
  \qp = \sum_i \epsilon_i \bigotimes_k^N \Lambda_{i_k},
\end{align}
where $\sum \epsilon_i = 1$. Substituting this expression into the partial trace in Eq.~\eqref{eqn:RedMat2}, we obtain (recall $\mathcal{S} = \{m,p\}$)
\begin{align}
  &\nonumber {\rm Tr}_{/\mathcal{S}}\left[\qp\left(\ketbra{\psi_{\mu_{\mathcal{S}}}}{\psi_{\nu_{\mathcal{S}}}}\otimes\mathbb{I}_{N-2}\right)\right] \\ =&~\nonumber2^{N-2}\sum_i \epsilon_i  \Lambda_{i_m}\otimes\Lambda_{i_p}\left(\ketbra{\psi_{\mu_{\mathcal{S}}}}{\psi_{\nu_{\mathcal{S}}}}\right) {\rm Tr}\left[\bigotimes_k^N \Lambda_{i_k}\left(\frac{\iden}{2}\right)\right]\\
  =&~\nonumber2^{N-2}\sum_i \epsilon_i \Lambda_{i_m}\otimes\Lambda_{i_p}\left(\ketbra{\psi_{\mu_{\mathcal{S}}}}{\psi_{\nu_{\mathcal{S}}}}\right)\\
  \equiv&~2^{N-2}\Lambda_{\mathcal{S}}\left(\ketbra{\psi_{\mu_{\mathcal{S}}}}{\psi_{\nu_{\mathcal{S}}}}\right),
\end{align}
where we have defined $\Lambda_{\mathcal{S}} = \sum_i\epsilon_i \Lambda_{i_m}\otimes\Lambda_{i_p}$. The reduced Choi state can then be written as
\begin{align}
  \rho_{\mathcal{S}}= \frac{1}{2^2}\sum_{\mu_{\mathcal{S}}\nu_{\mathcal{S}}}\ketbra{\psi_{\mu_{\mathcal{S}}}}{\psi_{\nu_{\mathcal{S}}}} \otimes\Lambda_{\mathcal{S}}\left(\ketbra{\psi_{\mu_{\mathcal{S}}}}{\psi_{\nu_{\mathcal{S}}}}\right), \label{eqn:RedMat3}
\end{align}
and it is clear that $\Lambda_{\mathcal{S}}$ must describe a valid quantum process.

In Eq.~\eqref{eqn:RedMat2} we see that the qubits outside the qubit pair of interest (the \emph{spectator} qubits) must be prepared in the maximally mixed state. If this is experimentally challenging, one can instead randomly sample spectator qubit preparations from the uniform distribution of the set of spectator qubit logical states. With sufficient sampling to generate accurate statistics, the normalized sum of the randomly sampled preparation states approaches the maximally mixed state for the spectator qubits. Thus, performing two-qubit QPT on the qubit pair of interest with spectator qubits prepared in a random logical state will characterize the desired effective process in Eq.~\eqref{eqn:RedMat2}.

From two-qubit QPT we can obtain an experimentally characterized two-qubit Choi state, which we label $\sigma_{\mathcal{S}}$. We equate this to our reduced Choi state for the unknown process, $\rho_{\mathcal{S}}$, to determine the free parameters in the PAPA. In other words, we simultaneously solve the equations
\begin{align}
  \rho_{\mathcal{S}} = \sigma_{\mathcal{S}}, \label{eqn:PAPAmain}
\end{align}
for every pair of qubits.

Note that each $\rho_{\mathcal{S}}$ depends on the $\chi$-matrix elements for \emph{all} qubit pairs, i.e.~all $\chi_{i_{k,n}}^{j_{k,n}}$, not just the qubit pair of the subset $\mathcal{S}$. Thus, each two-qubit process characterization $\sigma_{\mathcal{S}}$ constrains the global process, not just the component of the ansatz on the qubits in ${\mathcal{S}}$. For this reason, we have labelled the reduced two-qubit processes as $\Lambda_{\mathcal{S}}$ to distinguish them from the two-qubit processes that construct the PAPA, $\qp_{k,n+k}$ in Eq.~\eqref{eqn:2QPar}.

\subsection{PAPA and Gate Set Tomography}
\label{subsec:PAPAGST}

The PAPA tomography approach described so far works well to obtain a bootstrapped description of an $N$-qubit process from characterization of the effective processes on all qubit pairs. However, often the problem at hand is not to characterize a completely unknown process, but to determine the actual process, $\mathcal{G}$, that occurs when we aim to implement a unitary gate, $\hat{G}$, (from here on we use calligraphic text for processes and latin text for unitary gates).

Extending this to an entire gate set via Gate Set Tomography (GST), we obtain a set of processes $\{\mathcal{G}_i\}$ corresponding to the experimental implementation of an ideal gate set $\{\hat{G}_i\}$. GST has the further benefit of excluding state-preparation and measurement (SPAM) errors from the processes $\{\mathcal{G}_i\}$ \cite{Greenbaum:aa}. Note that for clarity we will use ``gate set'' to refer to the processes $\{\mathcal{G}_i\}$, and ``ideal gate set'' to refer to the unitary gates $\{\hat{G}_i\}$.

Combining PAPA with GST, we can perform GST on all qubit pairs to obtain a characterized gate set for each pair, and then use PAPA to bootstrap to descriptions of an $N$-qubit processes. To see why this is useful, consider the three-qubit gate $\hat{X}\otimes\hat{Y}\otimes\hat{X}$. Given characterized gate sets with the relevant two-qubit gates, one way to describe the three-qubit process would be
\begin{align}
  \hat{X}\otimes\hat{Y}\otimes\hat{X}\rho\hat{X}\otimes\hat{Y}\otimes\hat{X} \rightarrow \mathcal{G}_{X_1Y_2}\left(\mathcal{G}_{I_2X_3}\left(\rho\right)\right)
\end{align}
where $\mathcal{G}_{AB}$ is the experimental process when we try to implement the gate $\hat{A}\otimes\hat{B}$. However, there is ambiguity in the correct decomposition of the three-qubit gate, and $\mathcal{G}_{X_1X_3}\left(\mathcal{G}_{Y_2I_3}\left(\rho\right)\right)$ would be an equally valid description of the process. An issue arrises as it is unlikely that the constructed three-qubit processes from all possible two-qubit decompositions will agree with one another.

Using PAPA avoids this issue, as it finds the three-qubit process of PAPA form that best agrees with the pairwise characterized processes, i.e.~with $\mathcal{G}_{X_1Y_2}$, $\mathcal{G}_{Y_2X_3}$, and $\mathcal{G}_{X_1X_3}$. As such it captures context dependence between gate operations, such as when the effect on qubit 1 is different for the processes $\mathcal{G}_{X_1Y_2}$ and $\mathcal{G}_{X_1X_3}$. As an added benefit, one never has to implement the full $N$-qubit process, as one does when using PAPA without GST (as described in the previous section). Instead, from the characterized gate sets on all qubit pairs, we can bootstrap to PAPA characterizations of the processes in an $N$-qubit gate set, as represented in Fig.~\ref{fig:PAPA}b).

While PAPA can in principle return a characterization of any $N$-qubit gate, when we restrict the pairwise two-qubit QPT to GST, the PAPA+GST combination can only characterize a limited set of $N$-qubit gates. Which $N$-qubit gates can be characterized with PAPA+GST is detailed further in Appendix \ref{app:PAPAGST}. The general requirement is that each two-qubit reduced process of the ideal $N$-qubit gate must be an incoherent mixture of two-qubit gates built from the ideal gate set. For example, if the ideal gate is ${\rm CNOT}_{12}\otimes\hat{\iden}$, then as shown in Appendix \ref{app:PAPAGST}, the ideal gates $\hat{Z}\otimes\hat{\iden}$ and $\hat{\iden}\otimes\hat{\iden}$ need to be in the characterized gate set for qubit pair 1-3.

Decomposing an $N$-qubit gate this way implicitly assumes the errors that make the implemented process $\mathcal{G}$ distinct from the ideal gate $\hat{G}$ are not strongly specific to the implementation of $\mathcal{G}$. This is easily satisfied if the errors are gate-independent, but some kinds of gate-dependent error are tolerable, such as context dependence in simultaneous single-qubit gates. For the ${\rm CNOT}_{12}\otimes\hat{\iden}$ gate considered previously, an example of a tolerable gate-dependent error would be a coherent error that occurs on qubit 1 both for an actual $\hat{Z}$-gate or an effective $\hat{Z}$-gate (as occurs in the reduced process on qubit 1 for the ${\rm CNOT}_{12}$ gate).

It is important to emphasize that neither of these issues are limitations of PAPA, which can characterize any $N$-qubit process using pairwise two-qubit QPT, but of the two-qubit characterizations supplied to PAPA by GST. Nevertheless, there are many situations where PAPA+GST may be applicable, i.e.~the ideal-gate decomposition is possible and the errors can be assumed to be captured by PAPA+GST, as we explore in both experiment and theory, in sections \ref{sec:Exp} and \ref{sec:N12}.  For situations where PAPA+GST is not possible, PAPA can inherit SPAM-insensitivity from other SPAM-insensitive process tomography such as that using randomized benchmarking \cite{Kimmel:2014aa,Johnson:2015aa,Roth:2018aa}.

\section{Experimental Reconstructions}
\label{sec:Exp}
Here we test the PAPA+GST approach experimentally using an IBM five-qubit device similar to that of Ref.~\cite{Takita:2017aa}.  For this demonstration, we focus on a three-qubit subset of the chip with the goal of reconstructing three-body operations.  Device parameters, experimental diagrams and coherence times can be found in Appendix~\ref{app:device}.  To begin, two-qubit GST is performed on all three pairs of qubits in the subset, as is depicted in Fig.~\ref{fig:PAPA}b).  We choose the gate set $\{\hat{X}_{90}, \hat{Y}_{90}, \hat{\iden}\}^{\otimes 2}$ composed of simultaneous $90^{\circ}$ rotations around the $X$ and $Y$ axes, with the idle gate on both qubits (all $80~\mathrm{ns}$ long).  This set is chosen to allow the bootsrapping of non-trivial three-body operations and to avoid the issues discussed in the previous section~\ref{subsec:PAPAGST}.  In {\tt pyGSTi}, this gate set is called \verb|std2Q_XXYYII|.

\begin{figure}
  \centering
  \includegraphics[width=\columnwidth]{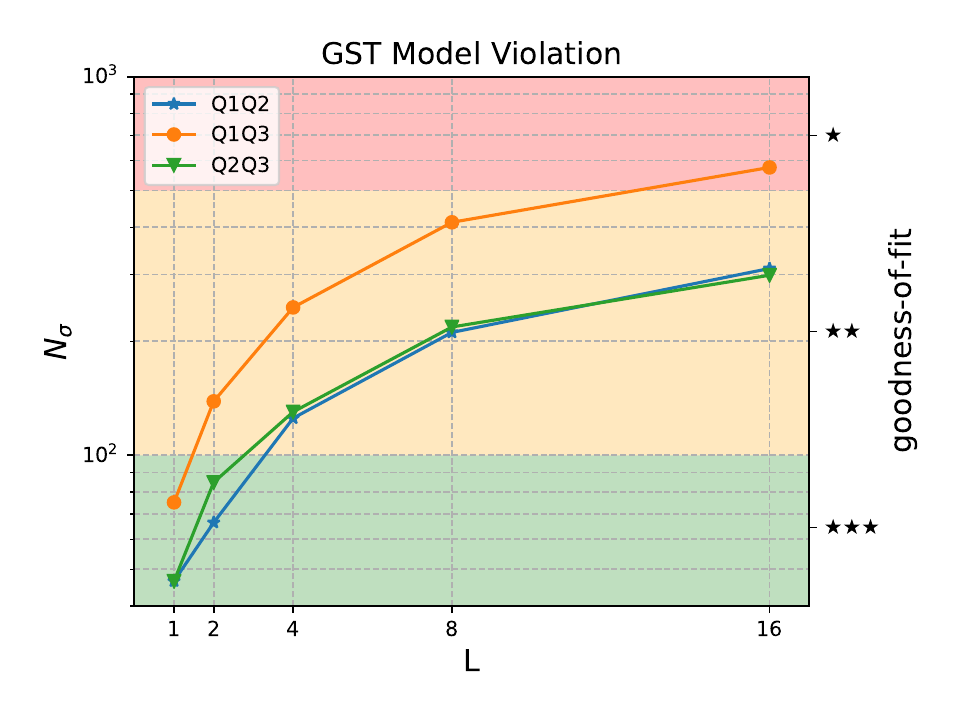}
  \caption{GST $N_\sigma$ vs. germ power $L$ for each of the three data sets.  $N_\sigma$ quantifies the deviation from a Markovian qubit model.  The goodness-of-fit parameter is provided by {\tt pyGSTi} and ranges from $\bigstar$ to $\bigstar\bigstar\bigstar\bigstar\bigstar$ indicating how well the data fits the implicit model.  Larger values of $L$ correspond to increased sensitivity to gate error and to longer circuits.}
  \label{fig:Nsigma}
\end{figure}

Gatestrings are generated with {\tt pyGSTi}~\cite{PYGSTI}, transpiled into our {\tt QGL}~\cite{QGL} language and finally compiled to a hardware specific format for our custom APS2 arbitrary waveform generators~\cite{Ryan:2017}.  To insure each GST experiment (across all three pairs) is subject to the same environmental noise on average, and as consistent as possible with other experiments, gatestrings from the three sets are interleaved on a shot-by-shot basis before being executed.  This prevents long term drift from changing system conditions across gatestrings, or pairs of qubits.  Additionally, to mimic the preparation of the spectator in the maximally mixed state (see Fig.~\ref{fig:PAPA}a)), each two-qubit GST experiment is repeated an equal number of times with the third qubit starting in either $\ket{0}$ or $\ket{1}$. The results are then combined and analyzed irrespective of the state of the spectator.  See Appendix~\ref{app:exp} for more details.

Experimental data is passed back to {\tt pyGSTi} for reconstruction.  Details of the reconstruction process can be found in~\cite{Blume-Kohout:2017aa} and~\cite{PYGSTI}.  To ensure viability of the PAPA process, the completely-positive trace-preserving (CPTP) constraint was enforced at every iteration $L$ as new data is added.  This guarantees a physical and consistent gate set is reconstructed.  The downside with this requirement is a considerable increase in runtime and RAM necessary for GST to converge.  To make this process tractable and time efficient, we used a Google Cloud instance~\cite{GOOGLE} with 40 vCPUs and 961 GB of RAM which allows all three GST data sets to be analyzed simultaneously.  The upper bound of 961 GB is not tight and was chosen out of an abundance of caution to ensure convergence.

Figure~\ref{fig:Nsigma} shows the $N_\sigma$ standard deviations from GST's implicit qubit model as a function of germ power $L$ for the three data sets.  The goodness-of-fit parameter on the right axes is supplied by {\tt pyGSTi} as a rough gauge for how well the model captures the dynamics in the data.  It is clear from the figure there are significant deviations at higher germ powers.  We attribute this to a combination of drift in system parameters, and leakage into higher excited states.

\begin{figure}
  \includegraphics[width=\columnwidth]{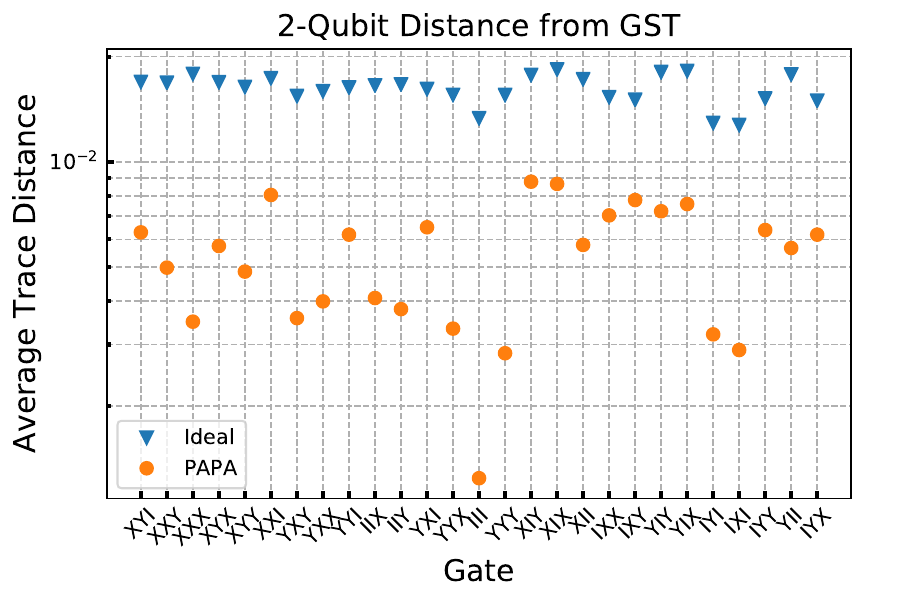}
  \caption{Comparison of the GST measured process matrices to the ideal and the PAPA reconstructions.  The PAPA data points are the average trace distance of the three reduced processes from there corresponding GST characterizations.  The ideal points are the average trace distance of the ideal gates to GST.  The trace distance (Eq.~\ref{eqn:2QTD}) for the PAPA reconstructions are lower for all 27 gates in the set.}
  \label{fig:PAPADIST}
\end{figure}

The final output of the GST algorithm yields three distinct local characterizations which are used by PAPA to reconstruct the larger three-qubit processes.  Each of the 27 three-qubit gates in the gate set are reconstructed simultaneously on a 32 core workstation using a \textsc{MATLAB} implementation of the PAPA algorithm.  Details of the non-linear least-squares bootstrapping at the heart of PAPA are outlined in Appendix~\ref{app:Num}.

The main experimental result of this letter is plotted in Fig.~\ref{fig:PAPADIST}.  For each three-qubit gate in the reconstructed gate set $\{\hat{X}_{90}, \hat{Y}_{90}, \hat{\iden}\}^{\otimes 3}$, we compare the GST characterizations of the effective two-qubit gates on each pair of qubits ($\sigma_{\mathcal{S}}$), to either the ideal reduced two-qubit gate, or the reduced two-qubit process obtained from the PAPA three-qubit reconstruction ($\rho_{\mathcal{S}}$). We quantify the distance between processes using the trace-distance
\begin{align}
  {\rm Trace~Dist.} = \frac{1}{2}{\rm Tr}\left[\sqrt{\left(\rho_{{\mathcal S}}-\sigma_{\mathcal S}\right)^\dagger\left(\rho_{{\mathcal S}}-\sigma_{\mathcal S}\right)}\right], \label{eqn:2QTD}
\end{align}
and the average of the three trace distances (one for each reduced two-qubit process) is what is plotted in Fig.~\ref{fig:PAPADIST}.

In all cases, PAPA produces an estimate of the process closer to the experimental data (GST characterizations) than the ideal gate.  Thus, from two-qubit tomography our PAPA+GST bootstrapping technique has produced a characterization of the three-qubit gate set that is both consistent with the tomography data (small trace distance in Fig.~\ref{fig:PAPADIST}), and consistent across pairs of qubits (by nature of the ansatz).

One may ask if PAPA is producing a characterization that could be explained with a simple model, such as single-qubit decoherence. However, a search over all possible values of $T_1$ and $T_2$ found no values that would make the data consistent with a simple decoherence model, and in fact such models did worse than the ideal gate.

\section{Simulation Tests of the Ansatz}
\label{sec:Sim}

\subsection{Noisy One- and Two-Qubit Gates}
\label{sec:N12}

To further test our PAPA approach for multi-qubit QPT, we numerically simulate ``unknown'' three-qubit processes, and then reconstruct the PAPA characterization of these processes. We consider several example processes formed by one of the ideal three-qubit gates $\hat{\iden}\otimes\hat{\iden}\otimes\hat{\iden}$, ${\rm CNOT}_{12}\otimes\hat{\mathbb{I}}$, or $\hat{X}\otimes\hat{Y}\otimes\hat{X}$, followed by an error process. For the error process we consider two cases of gate-independent error, either a coherent error described by single-qubit rotations on all three qubits
\begin{align}
  &\hat{G}_{\rm Coh.~Error} = \hat{X}_\phi\otimes\hat{Y}_\phi\otimes\hat{X}_\phi \\
  & \hat{X}_\phi = \cos(\phi)\hat{\iden} + i\sin(\phi)\hat{X} \\
  & \hat{Y}_\phi = \cos(\phi)\hat{\iden} + i\sin(\phi)\hat{Y}
\end{align}
or single-qubit decay and pure dephasing implemented by their standard Kraus operator representations \cite{Nielsen00}.

In standard PAPA reconstruction, pairwise two-qubit QPT is used to characterize the reduced two-qubit process, and obtain $\sigma_{\mathcal{S}}$ for each qubit pair. With PAPA+GST this is circumvented by using a GST characterized gate set for each qubit pair to calculate $\sigma_{\mathcal{S}}$, provided the ideal reduced two-qubit process can be built from gates in the ideal gate set. For the example three-qubit ideal gates chosen, the required two-qubit gates are contained in the ideal gate set ${\rm CNOT}+\{\hat{\mathbb{I}},\hat{X},\hat{Y},\hat{Z}\}^{\otimes 2}$. We follow the PAPA+GST approach for our numerical tests, simulating the implementation of this gate set on all qubit pairs, including the error process, and use results of these simulations as our GST reconstructed two-qubit gate sets. We then use the characterized two-qubit gate sets to calculate $\sigma_{\mathcal{S}}$ for each qubit pair. Explicit details of our approach are outlined in Appendix \ref{app:PAPAGST}.

\begin{figure}[!t]
  \begin{subfigure}{
  \includegraphics[width=0.8\columnwidth]{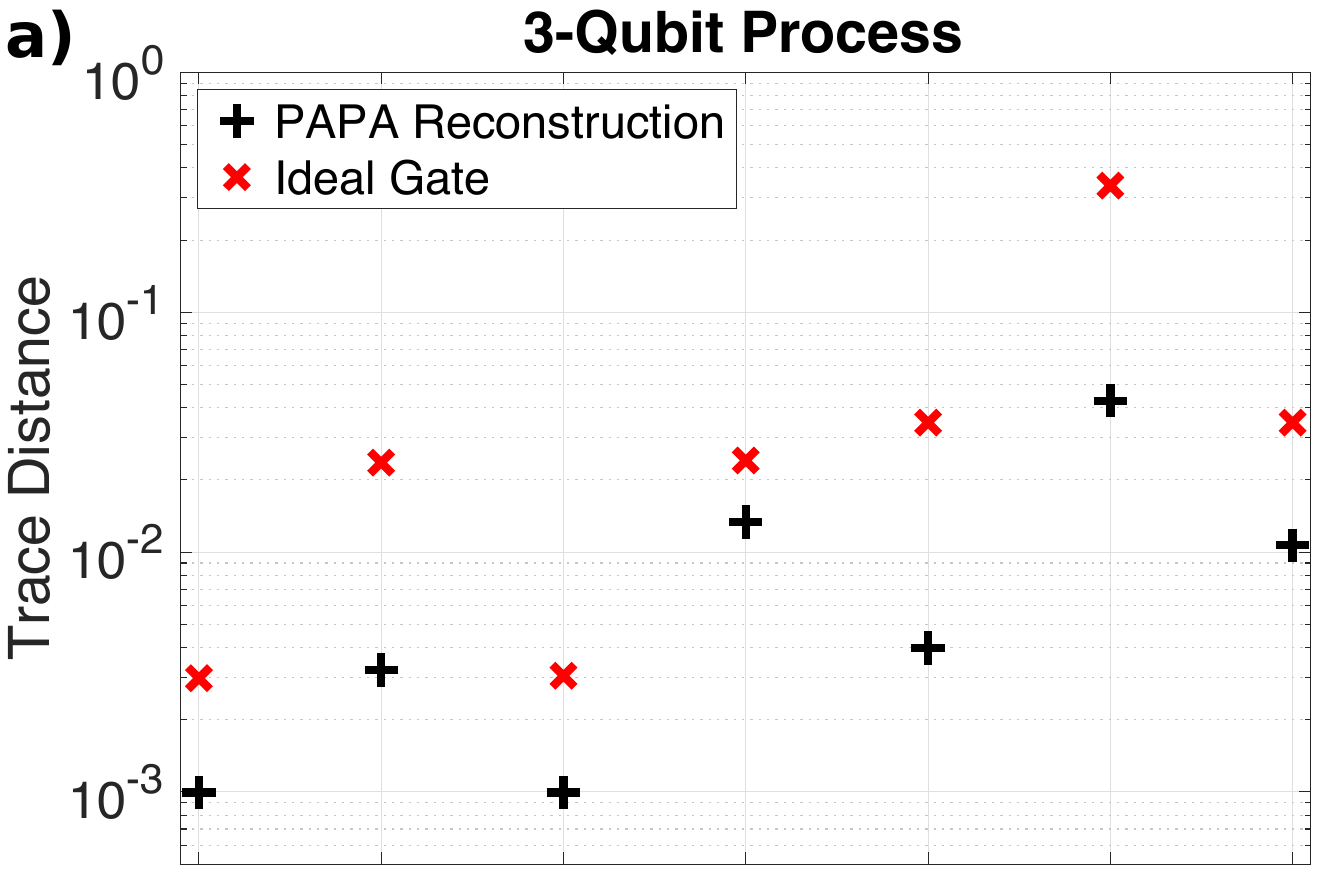}}\end{subfigure}
  \begin{subfigure}{
  \includegraphics[width=0.8\columnwidth]{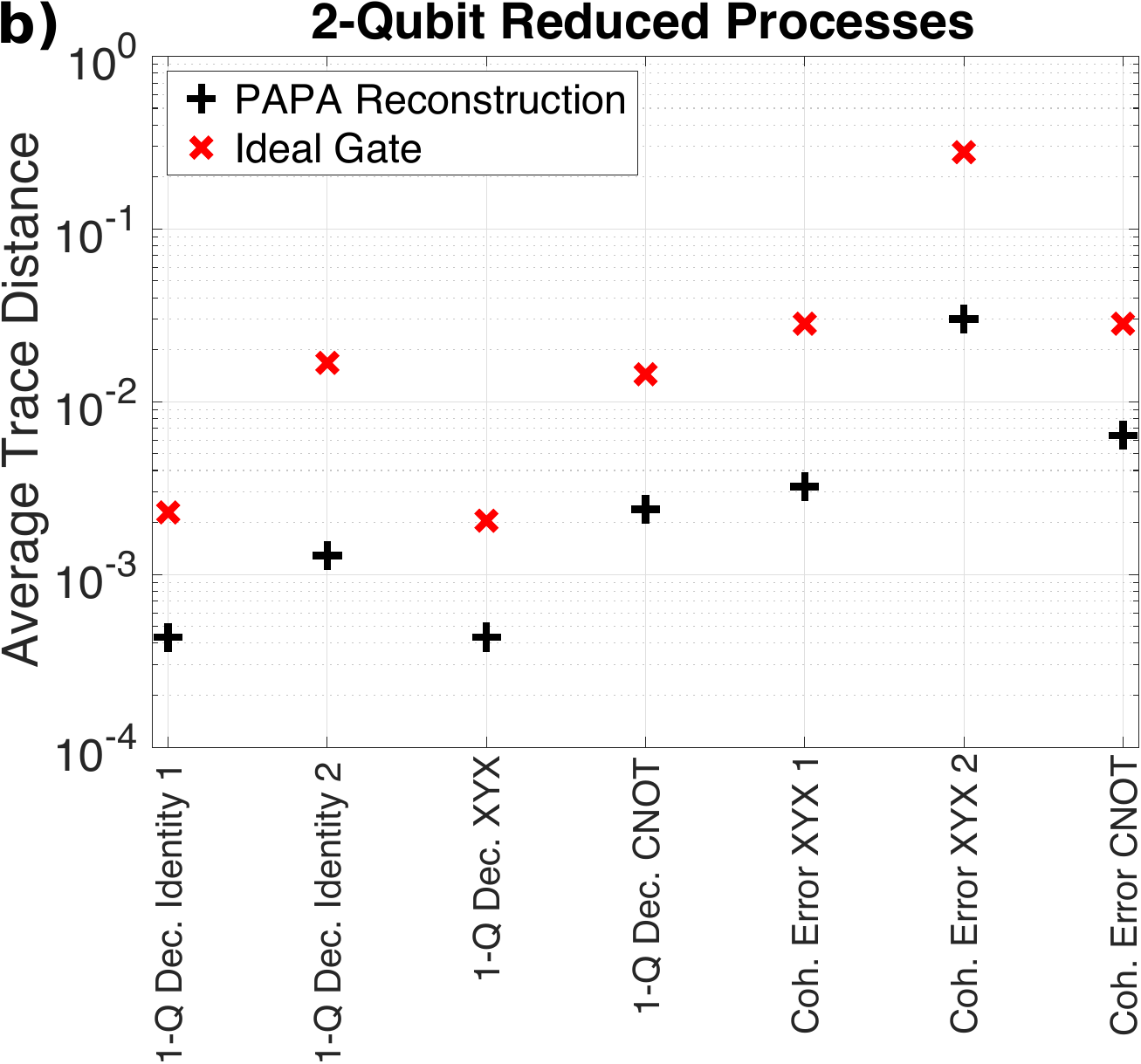}}\end{subfigure}
  \caption{{\bf a)} Simulated trace distance between the actual three-qubit process Choi state and either the PAPA reconstructed Choi state (black +) or the ideal gate (red $\times$), see Eq.~\eqref{eqn:3QTD}. {\bf b)} The average trace distance between the reduced two-qubit Choi states, see Eq.~\eqref{eqn:2QTD}. Processes i), ii) are the all identity gate of length 50 ns and 400 ns; iii), v), and vi) are $\hat{X}\otimes\hat{Y}\otimes\hat{X}$ of length 50 ns; iv) and vii) are ${\rm CNOT}_{12}\otimes\hat{\mathbb{I}}$ of length 400 ns. i), ii), iii), and iv) have single-qubit decoherence with $T_1 = T_2 = 50~\mu$s; v) and vii) have coherent error $\phi = 0.02$ and vi) has $\phi = 0.2$.}
  \label{fig:numsim}
\end{figure}

We compare the PAPA+GST reconstruction for a noisy gate to the actual simulated noisy gate by calculating the trace distance between the Choi state of the PAPA-reconstructed three-qubit process, $\rho_\qp$, and that for the actual process, $\rho^{\rm act}_\qp$
\begin{align}
  {\rm Trace~Dist.} = \frac{1}{2}{\rm Tr}\left[\sqrt{\left(\rho_\qp-\rho^{\rm act}_\qp\right)^\dagger\left(\rho_\qp-\rho^{\rm act}_\qp\right)}\right]. \label{eqn:3QTD}
\end{align}
In experiment we do not have access to the actual full three-qubit process, and cannot use the above trace distance as a performance metric. Instead, in section \ref{sec:Exp} we compared each of the reduced two-qubit processes from PAPA, to the actual two-qubit reduced processes from GST.  We do the same for our numerical tests.  The results of both trace distance calculations are shown in Fig.~\ref{fig:numsim} for the seven candidate processes listed in the caption.

As the results show, the PAPA reconstructed process always improves upon the initial guess (ideal gate), both in terms of the trace distance for the full three-qubit process reconstruction, Fig.~\ref{fig:numsim}a), and the average of the trace distances for the two-qubit reconstructions, Fig.~\ref{fig:numsim}b). This improvement is typically around one order of magnitude, except in the case of the CNOT gate, which was the most difficult to reconstruct of the gates tested.

The accuracy of the PAPA reconstructions for these simulated gates is set by the specifics of the classical numerical algorithm implemented (see Appendix \ref{app:Num} for details). If other algorithms \cite{Bolduc:2017aa,Knee:aa} more tailored to quantum process reconstruction are used with PAPA we expect significant improvements in accuracy and runtime are possible.

\subsection{Coherent Error in the Cross-Resonance Gate}

\begin{figure}[t]
  \includegraphics[width=\columnwidth]{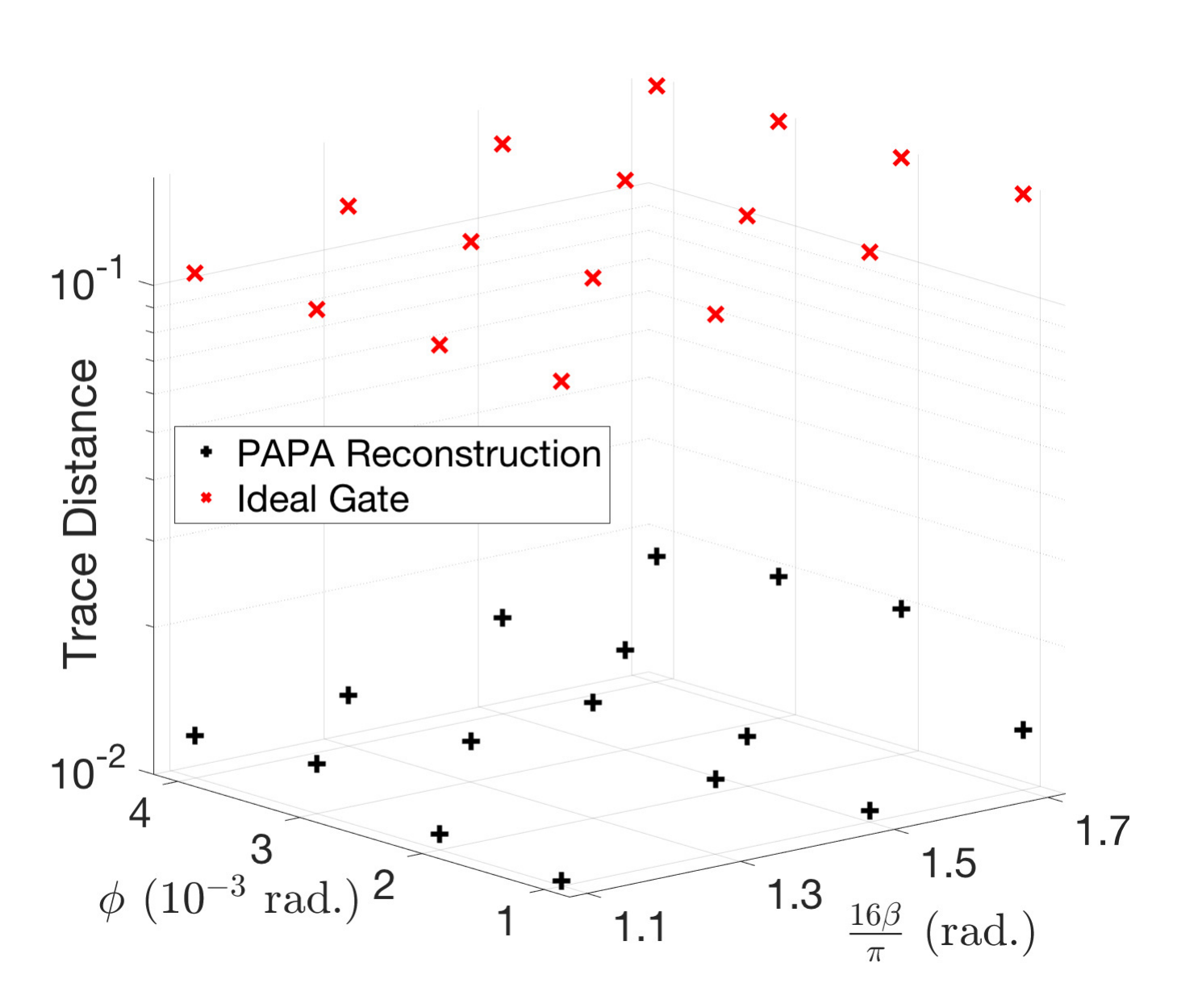}
  \caption{Simulated trace distance between the three-qubit Choi state for the simulated CR-CNOT with coherent error and either the PAPA reconstructed Choi state (black +), or the ideal gate (red $\times$), as a function of over-rotation error (angle $\beta$) and stray $ZZ$-coupling (angle $\phi$), see Eq.~\eqref{eqn:UCR}.}
  \label{fig:CR}
\end{figure}

We also perform a systematic testing of the PAPA approach by examining coherent error in a cross-resonance (CR) implementation of a CNOT gate \cite{Rigetti:2010aa,Chow:2011aa}, with the ideal gate taking the form ${\rm CNOT}_{12}\otimes \hat{\iden}$. Referred to as a CR-CNOT, this ideal gate consists of the ideal CR-gate followed by single-qubit gates. The unitary describing the implemented gate in the presence of coherent error is given by $\hat{U}_{\rm CNOT} = \hat{Z}^1_{-90}\hat{X}^{2}_{90}\hat{U}_{\rm CR}$, with the $\pm 90^{\circ}$ single-qubit rotations assumed to be perfect, and
\begin{align}
  \hat{U}_{\rm CR} = \exp\left(-i\left[\left(\frac{\pi}{2} + \beta\right)\frac{\hat{Z}\hat{X}\hat{\iden}}{2} + \phi~\frac{\hat{\iden}\hat{Z}\hat{Z}}{2}\right]\right), \label{eqn:UCR}
\end{align}
where for compactness of notation we have suppressed the tensor product symbols, such that $\hat{Z}\hat{X}\hat{\iden} = \hat{Z}\otimes\hat{X}\otimes\hat{\iden}$.

In Eq.~\eqref{eqn:UCR}, the angles $\beta$ and $\phi$ quantify the coherent error, with $\beta$ the angle of over-rotation from the desired CR-interaction between qubits 1-2, and $\phi$ the angle quantifying the effect of spurious $ZZ$-coupling between qubits 2-3. We consider the echoed CR-pulse of Ref.~\cite{Corcoles:2013aa}, such that the only remaining $ZZ$-coupling is between the target and idle qubits (i.e.~2 and 3). We use values of $\beta$ between $\pi/16$ and $\pi/8$ radians, which produce non-ideal gates with trace-overlap fidelity of $95-99\%$, and values of $\phi$ between $10^{-3}$ and $4\times 10^{-3}$ radians. For a gate of 400 ns in duration, these values of $\phi$ correspond to spurious ZZ-couplings of $2.5-10$ kHz.

The error introduced to the CR-CNOT in Eq.~\eqref{eqn:UCR} is strongly gate-dependent, since it is intrinsic to the CR interaction itself. However, the reduced two-qubit gate decomposition of the CNOT (see Appendix \ref{app:PAPAGST}) used in PAPA+GST contains gates that do not involve the CR interaction. These gates will be insensitive to the CR error, and as a result PAPA+GST is not applicable in this situation. Instead we apply standard PAPA, and simulate QPT on the effective process for each pair of qubits during the implemented CR-CNOT. For this we assume no SPAM error, and in practice similar results can be achieved by applying other SPAM-insensitive process tomography approaches to the CR-CNOT \cite{Kimmel:2014aa,Johnson:2015aa,Roth:2018aa}.

The results of our simulations are shown in Fig.~\ref{fig:CR}. As can be seen, for all values of $\beta$ and $\phi$ tested the PAPA reconstruction is approximately an order of magnitude closer to the simulated unitary of Eq.~\eqref{eqn:UCR} than the ideal gate (used as the initial guess). Thus, PAPA is a useful technique for benchmarking the performance of experimentally relevant implementations of entangling gates, such as the CR-CNOT widely used in circuit QED \cite{Chow:2014aa}.

\section{Conclusion}
\label{sec:Conc}

We have presented here an approach to efficient and SPAM-insensitive quantum process tomography that relies on fitting tomographic data to a constrained ansatz for the unknown quantum process. Our physically motivated pairwise perturbative ansatz requires only two-qubit process tomography on all pairs of qubits, such that the total number of measurements scales only quadratically with qubit number. Further, our ansatz inherits SPAM-insensitivity from SPAM-insensitive two-qubit tomography, such as gate set tomography \cite{Blume-Kohout:2017aa} or RB gate tomography \cite{Kimmel:2014aa,Johnson:2015aa,Roth:2018aa}.

The experimental demonstration of PAPA shows a significant improvement in the accuracy of reconstructed two-qubit processes calculated from the bootstrapped three-qubit process.  Testing via numerical simulations validates the usefulness of our tomographic approach on both a series of example gates, and the experimentally relevant CR-CNOT \cite{Chow:2011aa}. In typical cases, the resulting description of the unknown quantum process found by our ansatz is an order of magnitude more accurate than the naïve initial guess. In the future, we hope to improve the efficiency and accuracy of the classical algorithm underlying our reconstruction method \cite{Bolduc:2017aa,Knee:aa}.

It is worth noting that while we have chosen to build our ansatz for an $N$-qubit process from two-qubit processes, similar ansatz can be created from $K$-qubit processes for any $K < N$. These have measurement resource requirements that scale as a polynomial of order $K$, and are therefore still asymptotically efficient. We have focussed on the case $K=2$ in our work as two-qubit process tomography is within current experimental capabilities. However, for larger system sizes, there will likely be an optimal $K>2$ that reduces the number of qubit subsets, given by $N\choose K$, while maintaining a small enough $K$ that $K$-qubit QPT is experimentally feasible.

Finally, we comment briefly on the situations where PAPA may fail, and the fact that this actually gives useful information about the unknown process. Numerical reasons aside, PAPA reconstruction fails when the process being estimated is an operation that is not factorable to 2-body, or when non-Markovian noise is present. As such, PAPA reconstruction can be used as a form of model testing for error processes that entangle more than 2 qubits, or non-Markovian noise sources such as slow parameter drift. Similarly, PAPA+GST puts greater restrictions on the gate and context independence of the noise sources, and can be used as a model testing procedure for these error sources. This highlights the usefulness of ansatz-based approaches to QPT: even when they fail they provide useful information about the system.

\acknowledgements

The authors acknowledge useful discussions with David Poulin. The qubit device and amplifiers used in the experiment were graciously provided by IBM under the Intelligence Advanced Research Projects Activity (IARPA) contract W911NF-16-0114.  This material is based upon work supported by the U.S. Army Research Office under Contract No: W911NF-14-C-0048. Any opinions, findings and conclusions or recommendations expressed in this material are those of the authors and do not necessarily reflect the views of the U.S. Army Research Office.

\appendix

\section{PAPA+GST $N$-qubit Gates}
\label{app:PAPAGST}

In this appendix we discuss the set of $N$-qubit gates that can be characterized via bootstrapping with PAPA+GST. We will focus on the $N=3$ case since the extension to $N>3$ is straightforward from the three-qubit results. Consider an ideal three-qubit gate written as
\begin{align}
 \hat{U} = \sum_{ijk} u_{ijk} \hat{U}_{i}\otimes\hat{U}_{j}\otimes\hat{U}_{k},
\end{align}
where ${\rm Tr}(\hat{U}_i\hat{U}^\dagger_j) = 2\delta_{ij}$ such that $\{\hat{U}_i\}$ is an orthonormal basis for one-qubit operator space. We will label the ideal process for this gate as $\mathcal{U}$, and label the imperfect experimental implementation of this process as $\tilde{\mathcal{U}}$. For notational simplicity we break slightly from the nomenclature used in the main text, and throughout this appendix processes without tildes will be ideal, and those with tildes will be experimental implementations of the ideal process.

The Choi state of the ideal process is
\begin{align}
  &\rho_{\mathcal{U}} = \frac{1}{8}\sum_{\mu\nu}\ketbra{\psi_\mu}{\psi_\nu} \\
  &\nonumber\otimes \sum_{\substack{ijk \\i'j'k'}} u_{ijk}u_{i'j'k'} \hat{U}_{i}\otimes\hat{U}_{j}\otimes\hat{U}_{k}\ketbra{\psi_\mu}{\psi_\nu} \hat{U}_{i'}^\dagger\otimes\hat{U}_{j'}^\dagger\otimes\hat{U}_{k'}^\dagger.
\end{align}
Then, as an example, the two-qubit reduction for qubits 1-2 is given by
\begin{align}
  \nonumber\rho_{\mathcal{U}_{1,2}} &= \frac{1}{4}\sum_{\mu\nu}\ketbra{\psi_\mu}{\psi_\nu} \\
  &\nonumber\otimes \sum_{\substack{ij \\i'j'}}\sum_k u_{ijk}u_{i'j'k} \hat{U}_{i}\otimes\hat{U}_{j}\ketbra{\psi_\mu}{\psi_\nu} \hat{U}_{i'}^\dagger\otimes\hat{U}_{j'}^\dagger \\
  &= \frac{1}{4}\sum_{\mu\nu}\ketbra{\psi_\mu}{\psi_\nu}\otimes\mathcal{U}_{12}(\ketbra{\psi_\mu}{\psi_\nu}),
\end{align}
where we have used the fact that ${\rm Tr}(\hat{U}_i\hat{\iden}\hat{U}^\dagger_j) = 2\delta_{ij}$, and $\mathcal{U}_{12}$ is the two-qubit process defined by
\begin{align}
  \mathcal{U}_{12}(\rho) = \sum_k u_{ijk}u_{i'j'k} \hat{U}_{i}\otimes\hat{U}_{j}\rho \hat{U}_{i'}^\dagger\otimes\hat{U}_{j'}^\dagger.
\end{align}

The general PAPA approach would be to characterize the experimental implementation of the process $\mathcal{U}_{12}$, i.e.~$\tilde{\mathcal{U}}_{12}$, via two-qubit QPT on qubits 1-2 when $\tilde{\mathcal{U}}$ occurs. The PAPA+GST approach is the situation where one \emph{does not} want to perform two-qubit QPT for every unknown three-qubit process, but would rather bootstrap characterizations of three-qubit processes from existing two-qubit gate set characterizations.

In the PAPA+GST approach, the two-qubit reduction $\rho_{\tilde{\mathcal{U}}_{1,2}}$ can be experimentally characterized if the ideal process $\mathcal{U}_{12}$ can be described as a convex sum of unitary processes
\begin{align}
  \mathcal{U}_{12}(\rho) = \sum_i c_i \hat{G}_i\rho\hat{G}_i^\dagger,
\end{align}
with each $\hat{G}_i$ in the GST characterized gate set. If this is the case, then
\begin{align}
  \rho_{\tilde{\mathcal{U}}_{1,2}} = \sum_i c_i \sigma_{\tilde{\mathcal{G}}_i},
\end{align}
where $\tilde{\mathcal{G}}_i$ is the experimental implementation of the gate $\hat{G}_i$, and each $\sigma_{\tilde{\mathcal{G}}_i}$ can be obtained from the GST gate set which contains all $\tilde{\mathcal{G}}_i$.

For the ideal gate set we have used in the main text, ${\rm CNOT}+\{\hat{\mathbb{I}},\hat{X},\hat{Y},\hat{Z}\}^{\otimes 2}$, we will now show that any three-qubit quantum gate consisting of a single-layer circuit of these gates can be characterized using PAPA+GST. Three-qubit gates of the form $\hat{G}_1\otimes\hat{G}_2\otimes\hat{G}_3 \in \{\hat{\mathbb{I}},\hat{X},\hat{Y},\hat{Z}\}^{\otimes 3}$ can obviously be parameterized by PAPA, as one can trivially show that the two-qubit processes to be characterized are the unitary gates $\hat{G}_1\otimes\hat{G}_2$, $\hat{G}_1\otimes\hat{G}_3$, and $\hat{G}_2\otimes\hat{G}_3$, which are all in the GST gate sets.

For a three-qubit gate that involves a CNOT on two of the qubits, a bit more effort is required to show that the necessary two-qubit gates to be characterized are still in the GST gate set. For example, consider the ideal gate $\hat{U} = {\rm CNOT}_{12}\otimes\hat{\iden}$ used in the main text. This has the ideal two-qubit reduced dual states
\begin{align}
  &\rho_{\mathcal{U}_{1,2}} = \frac{1}{4}\sum_{\mu\nu}\ketbra{\psi_\mu}{\psi_\nu}\otimes{\rm CNOT}\ketbra{\psi_\mu}{\psi_\nu}{\rm CNOT} \label{eqn:Cx1} \\
  &\rho_{\mathcal{U}_{1,3}} = \frac{1}{2}\left(\rho_{\iden\otimes\iden} + \rho_{Z\otimes\iden}\right) \label{eqn:Cx2} \\
    &\rho_{\mathcal{U}_{2,3}} = \frac{1}{2}\left(\rho_{\iden\otimes\iden} + \rho_{X\otimes\iden}\right) \label{eqn:Cx3}
\end{align}
and therefore the necessary gates to characterize are ${\rm CNOT}$ for qubits 1-2, $\hat{\iden}\otimes\hat{\iden}$ and $\hat{Z}\otimes\hat{\iden}$ for qubits 1-3, $\hat{\iden}\otimes\hat{\iden}$ and $\hat{X}\otimes\hat{\iden}$ for qubits 2-3. As all of these gates belong to their respective GST gate sets, a characterization of $\hat{U}$ can be bootstrapped using PAPA+GST. It is straightforward to show that this generalizes to all arrangements of the CNOT (i.e.~on any pair of qubits), and any gate on the qubit not involved in the CNOT.

So far we have only commented on the ideal two-qubit gates that need to be characterized for a given $N$-qubit process, and not on the other criteria for PAPA+GST, namely tolerable error. The general criteria is not as strong as all error needing to be gate-independent. For instance, three-qubit gates such as $\hat{G}_1\otimes\hat{G}_2\otimes\hat{G}_3$ may have error that is dependent on the specific single-qubit gates implemented, as this will be captured in characterization of the two-qubit reductions. Similarly, if the error in a single-qubit gate depends on the gate implemented on another qubit (i.e. context dependence) this will also be captured by PAPA+GST.

The fact that both gate-dependent and context-dependent error fits within the PAPA+GST framework for simultaneous single-qubit gates comes from the fact that the physical implementation of the simultaneous single-qubit gates on $N$ qubits is the same as on two-qubits. This is often not the case for an entangling gate such as ${\rm CNOT}_{12}\otimes\hat{\iden}$, where the physical implementation, such as a CR-CNOT, could be vastly different than the physical implementation of the gates in its reduced two-qubit decomposition, cf. Eqs.~\eqref{eqn:Cx1}-\eqref{eqn:Cx3}.

This is especially true in the case of the CR-CNOT where the error model assumed in Eq.~\eqref{eqn:UCR} -- over-rotation plus cross-talk -- is intrinsic to the CR interaction.  As such, GST characterization of the simultaneous single-qubit gates $\hat{Z}\otimes\hat{\iden}$ and $\hat{X}\otimes\hat{\iden}$ (on qubit pairs 1-3 and 2-3 respectively) would not contain any signature of this error. This makes PAPA+GST impossible, as two qubit tomography from GST would be inconsistent across qubit pairs.  Even a large difference in gate-length between entangling and single-qubit gates can result in an error discrepancy due to decoherence, and this is enough to make PAPA+GST inapplicable. In such situations standard PAPA should be used in combination with other SPAM-insensitive two-qubit QPT techniques.

\section{Device Parameters}
\label{app:device}
Table~\ref{tab:DP} shows device performance for all five qubits on the sample.  Note for the experiment presented only $Q_1$, $Q_2$ and $Q_3$ were used.  All gate times are $80~\mathrm{ns}$.
\begin{table}[h!]
\begingroup
\setlength{\tabcolsep}{5pt} 
\renewcommand{\arraystretch}{1.2} 
\begin{tabular}{@{}ccccccccccccccccc@{}}
\toprule
Qubit    & $f_{01}$ (GHz) & $T_2^\ast$ ($\mu$s)  & $T_1$ ($\mu$s) & $ZZ~\mathrm{(kHz)}$ \\
\midrule
$Q_1$      & 5.3067         & $21 \pm 5$                     & $43 \pm 9$            & \begin{tabular}{cc} 30 ($Z_1Z_4$);\\ 25 ($Z_1Z_2$) \end{tabular} \\
$Q_2$       & 5.2125         & $11 \pm 3$                    & $8.3 \pm 0.3$            &  8 ($Z_2Z_3$) \\
$Q_3$    & 5.357          & $43 \pm 9$                     & $50 \pm 6$        &  35 ($Z_3Z_5$) \\
$Q_4$       & 5.4177         &  $20 \pm 7$                  &  $44 \pm 6$ &  32 ($Z_2Z_4$) \\
$Q_5$       & 5.4123         & $22 \pm 6$                 &  $45 \pm 3$ & 72 ($Z_2Z_5$)               \\
\bottomrule
\end{tabular}
\endgroup
\caption{Qubit frequencies and coherence times. $Q_4$ and $Q_5$ were not used in this experiment. Coherence times are quoted as average and standard deviation of the values measured over a $\sim 16~\mathrm{h}$ span. A window of $2~\mathrm{h}$ was not included in the statistics for $Q_1$, during which $T_1$ was suppressed most likely due to a TLS moving into resonance with the qubit~\cite{Klimov:2018}.  Static $ZZ$ interaction strengths are reported for qubit pairs that are coupled through a common bus resonator.}
\label{tab:DP}
\end{table}

\section{Experimental Methods}
\label{app:exp}

\begin{center}
\begin{figure}[t]
  \includegraphics[width=1.0\columnwidth]{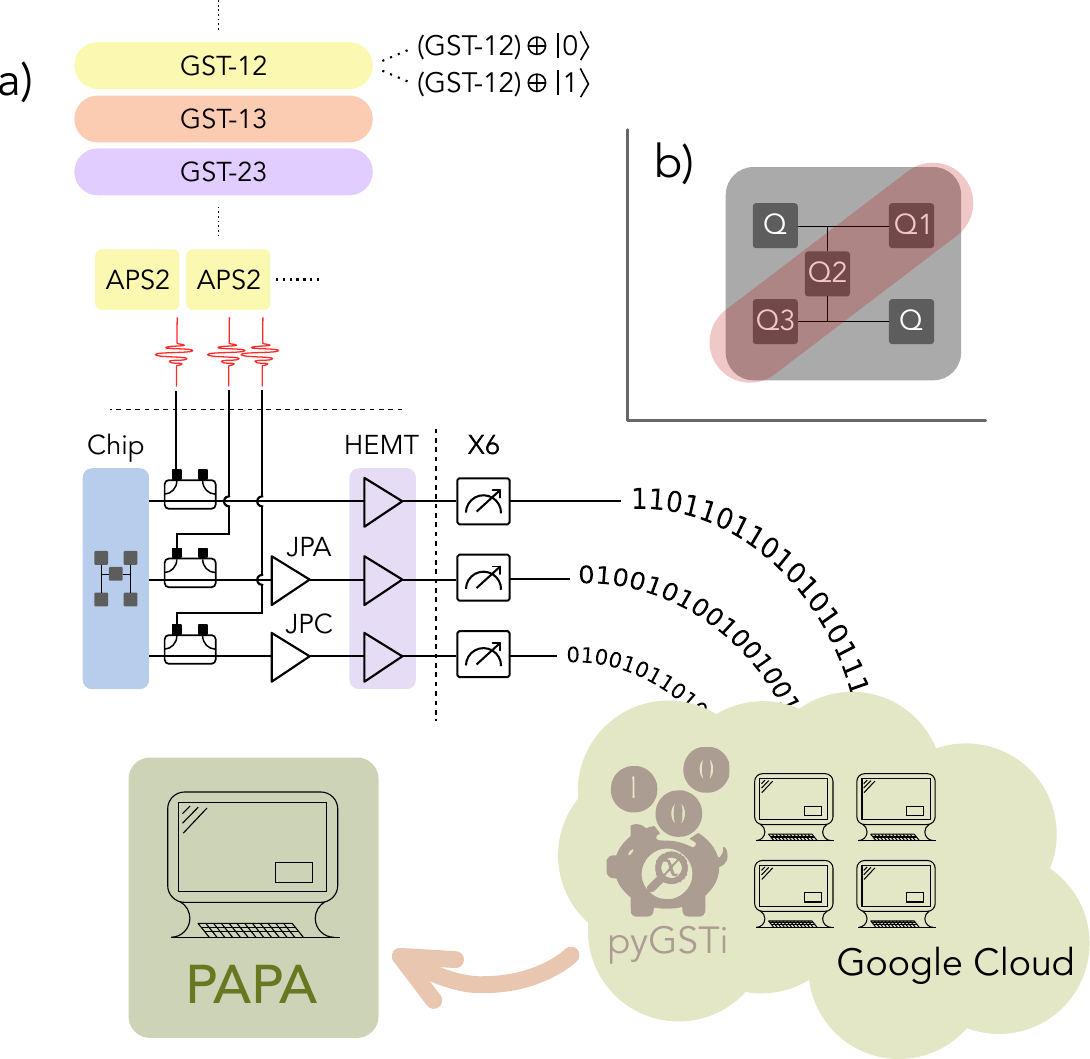}
  \caption{PAPA experimental data flow.  {\bf a)} GST experiments are interleaved by pair and spectator qubit state and sent to the BBN Arbitrary Pulse Sequencers (APS2)~\cite{Ryan:2017}.  Gate instructions are converted to control pulses and sent to the five-qubit device.  Demodulated readout signals are first amplified by HEMTs, JPAs and JPCs then digitized into qubit state information with custom firmware running on an X6-1000M digitizer card.  This data is then passed to {\tt pyGSTi} and reconstructed in parallel using Google Cloud Compute services.  The three separate reconstructions are then passed to the PAPA algorithm for bootstrapping to three-qubit processes. {\bf b)} A notional diagram of the IBM five-qubit device used in the experiment.  The location of the $\{Q1, Q2, Q3\}$ subset is specified in red.  Lines denote static capacitive coupling through CPW resonators.}
  \label{fig:diagram}
\end{figure}
\end{center}

Qubit control and readout are performed through dedicated co-plainer waveguide (CPW) resonators, one coupled to each qubit~\cite{Takita:2017aa}. $Q_1$ is equipped with a Josephson parametric amplifier from UC Berkeley~\cite{Hatridge:2011} and $Q_3$ with a Josephson parametric converter from IBM~\cite{Abdo:2013} for improved readout fidelity. All the pulses are generated using the BBN pulse generators introduced in Ref.~\cite{Ryan:2017}. Readout signals are acquired and processed using 2 Innovative Integration X6-1000 digitizers programmed with the BBN QDSP firmware ({\it ibid}). In particular, for each measured qubit, a $2.3~\mu$s homodyne signal is integrated using a pre-calibrated matched filter~\cite{Ryan:2015}, and subsequently reduced to a single-bit value according to an optimized threshold. All of this signal processing takes place on the digitizer FPGA board, thus expediting data acquisition and writing to disk. The results, which are digitized independently for each qubit, are then converted into a number of counts for each of the 4 combinations in a qubit pair, which is the input format for pyGSTi.  This process is illustrated in Fig.~\ref{fig:diagram}.

The 27 three-qubit gates were bootstrapped with PAPA on a 32 vCPU workstation with 32 GB of RAM.  The \textsc{MATLAB} implementation of of PAPA took roughly 24 hours to complete for low-weight gates like $\hat \iden \otimes \hat \iden \otimes \hat \iden$ and close to 48 hours for high-weight gates like $\hat X \otimes \hat X \otimes \hat X$.  This was done using a single core and less than a GB of RAM per gate.  We stress again, this implementation of PAPA can be much improved in terms of time and resources required.  A more optimized version of the code is currently under development which we plan to open source to the community.

\section{Numerical Implementation}
\label{app:Num}

\begin{figure}[t]
  \includegraphics[width=\columnwidth]{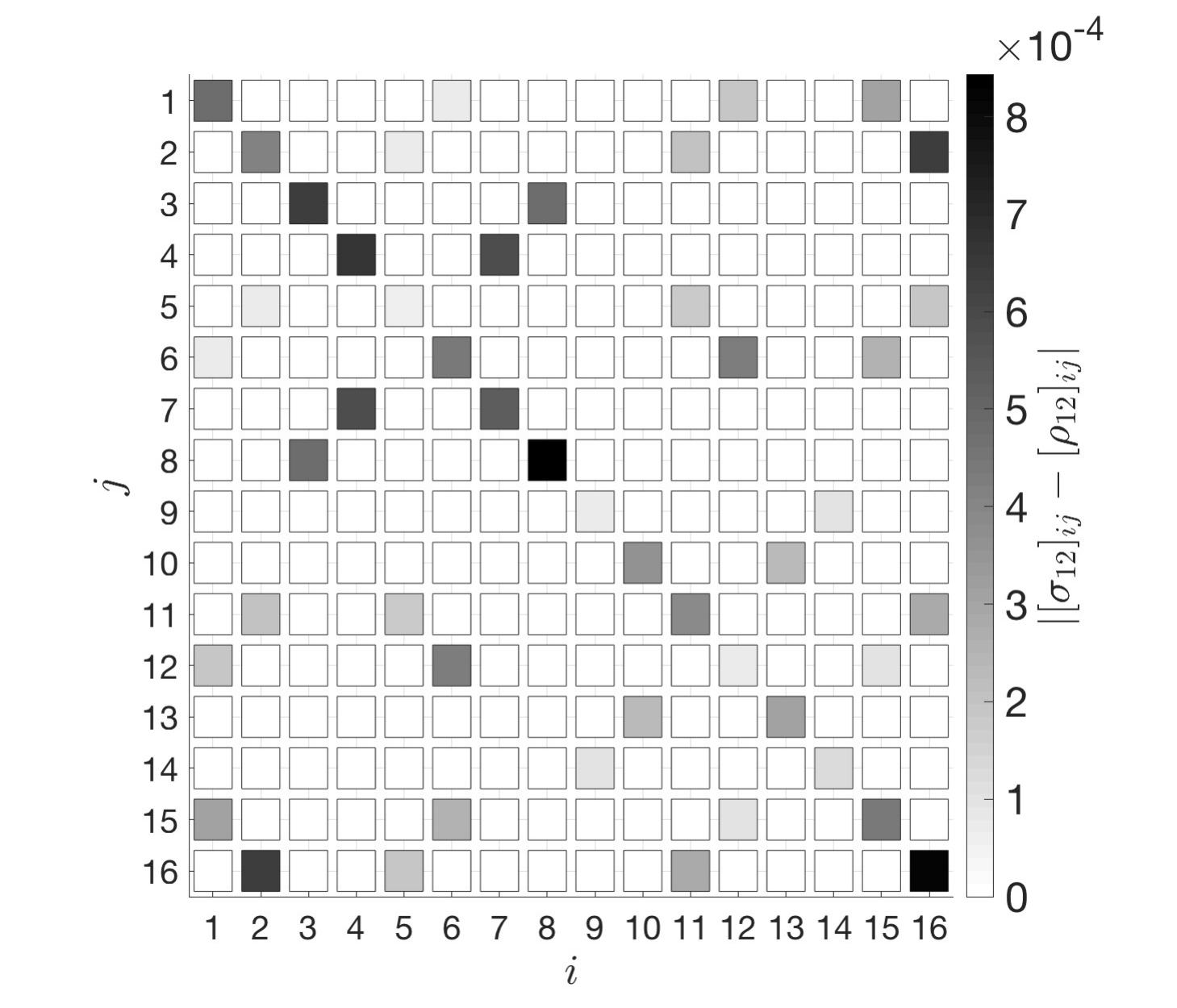}
  \caption{Absolute value of the element-wise difference between the ``measured'' Choi state, $\sigma_{12}$, and the PAPA reconstructed Choi state, $\rho_{12}$, for the effective process experienced by qubit pair 1-2 during a CNOT gate with single-qubit decoherence. Simulation parameters for the three-qubit process are the same as in Fig.~\ref{fig:numsim}.}
  \label{fig:CS}
\end{figure}

The computational task in PAPA characterization is the simultaneous solution of Eq.~\eqref{eqn:PAPAmain} for each pairwise reduction, from which we obtain the elements of the two-qubit $\chi$-matrices, $\chi^{j_{k,n}}_{i_{k,n}}$. These equations are nonlinear in general, and must be solved under the constraint that each of the two-qubit $\chi$-matrices describes a completely-positive and trace-persevering (CPTP) map.

This implies that the $\chi$-matrix is a positive semi-definite matrix with trace 4 (dimension of two-qubit Hilbert space). Further, CP requires an additional constraint, which to describe we need to parameterize a two-qubit process on the set $\mathcal{S}$ in the usual way via its $\chi$-matrix
\begin{align}
  \qp_{\mathcal{S}}(\rho) = \sum_{p,r}^{16} [\bm{\chi}_{\mathcal{S}}]_{p,r}\hat{E}_r\rho\hat{E}_p^\dagger,
\end{align}
where $\{\hat{E}_p\}$ is a basis for two-qubit operator space. The CP constraint is then \cite{Nielsen00}
\begin{align}
  \sum_{p,r}^{16} [\bm{\chi}_{\mathcal{S}}]_{p,r}\hat{E}_p^\dagger\hat{E}_r - \hat{\mathbb{I}}\otimes\hat{\mathbb{I}} = 0. \label{eqn:CPcon}
\end{align}

\begin{figure}[t]
  \includegraphics[width=0.9\columnwidth]{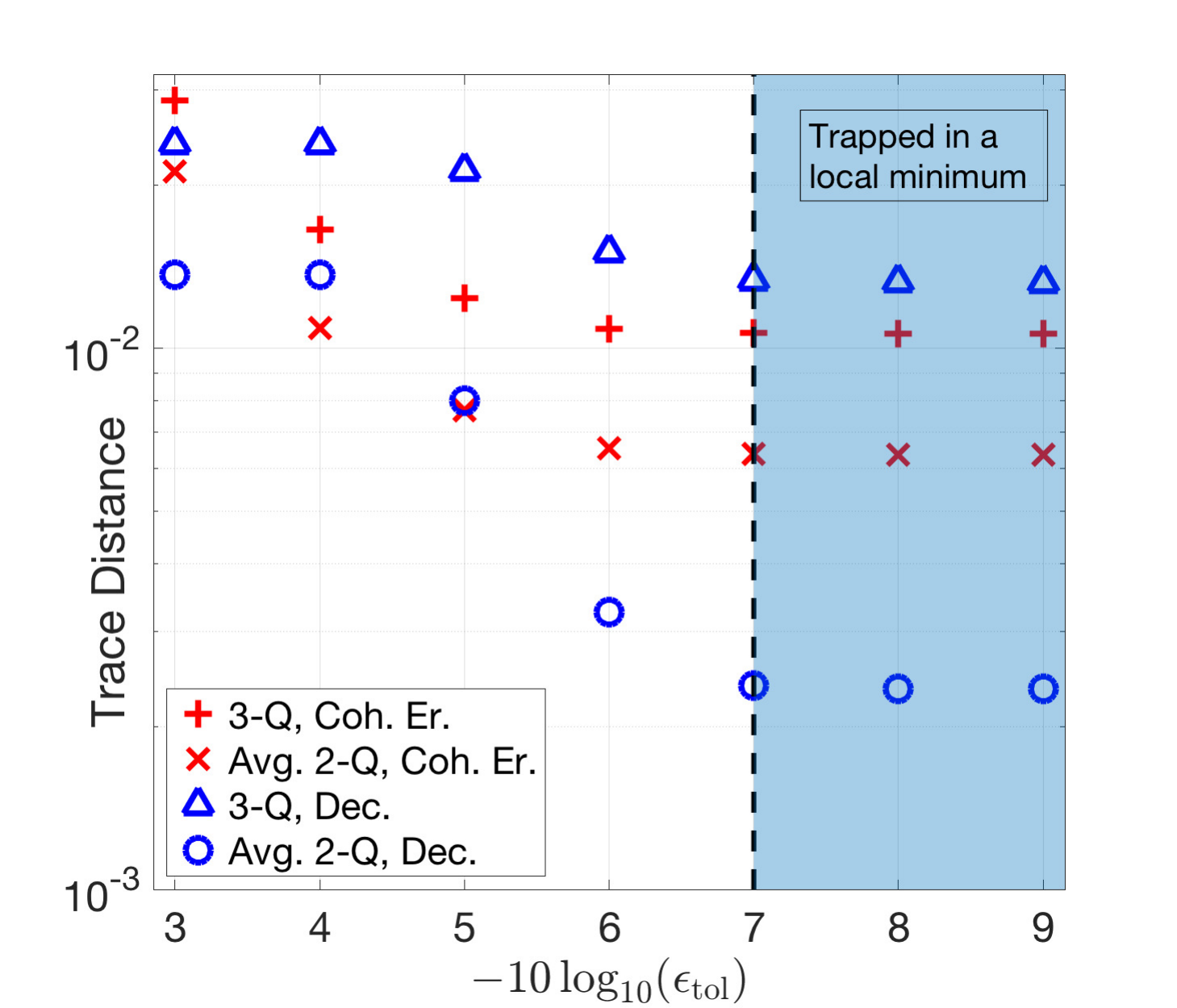}
  \caption{Trace distance between the simulated Choi state and PAPA reconstructed Choi state for the three-qubit process, and reduced two-qubit processes as a function of the solver tolerance, $\epsilon_{\rm tol}$. The CNOT gate with coherent error (red $+$ and $\times$) and decoherence (blue triangles and circles) were used for these simulations. The vertical dashed line indicates the tolerance used for the simulations presented in the main text.}
  \label{fig:tol}
\end{figure}

Comparing this to our previous parameterization of a two-qubit process in terms of one-qubit processes used in Eq.~\eqref{eqn:2Qlocal},
\begin{align}
  \qp_{\mathcal{S}} = \sum^{16}_{i,j_{k,n}}\chi_{i_{k,n}}^{j_{k,n}}\mathcal{A}^{(k)}_{i_{k,n}}\otimes\mathcal{A}^{(k+n)}_{j_{k,n}},
\end{align}
for $\mathcal{S} = \{k,k+n\}$, we see that the two parameterizations are related by splitting each index $i_{k,n}$ and $j_{k,n}$ into two parts via the equations
\begin{align}
  &\nonumber\mathcal{A}^{(k)}_{i_{k,n}}\otimes\mathcal{A}^{(k+n)}_{j_{k,n}}(\rho) = \hat{E}_{r_1}\otimes\hat{E}_{r_2}\rho\hat{E}_{p_1}^\dagger \otimes \hat{E}_{p_2}^\dagger = \hat{E}_r\rho\hat{E}_p^\dagger, \\
  &\nonumber i_{k,n} \rightarrow (i, i')~~~j_{k,n} \rightarrow j, j', \\
  &\nonumber r = (r_1,r_2) = (i,j)~~~p = (p_1,p_2) = (i',j'), \\
  &\nonumber [\bm{\chi}_{\mathcal{S}}]_{p,r} = \chi_{(r_1,p_1)}^{(r_2,p_2)}.
\end{align}

To solve for the $\chi$-matrix elements under the CPTP constraints, we use a least-squares minimization approach implemented in \textsc{MATLAB} \cite{mathworks}. Here, the cost function for the least-squares minimization consists of two parts. The first encodes the experimental characterizations of the two-qubit reductions, and simply consists of the element-wise difference between the two-qubit reduced Choi state for each pair of qubits and the current estimate for the two-qubit reduced Choi state generated by PAPA,
\begin{align}
  C_1\left[\vec{\bm{\chi}}\right] = \sum_{\mathcal S}\sum_{k,n}\abs{\left[\rho_{\mathcal{S}}(\vec{\bm{\chi}})\right]_{k,n} - \left[\sigma_{\mathcal{S}}\right]_{k,n}}^2,
\end{align}
where $\vec{\bm{\chi}}$ is a vector of the $\chi$-matrices for the processes on all qubit pairs that make up the PAPA, and the sum over $\mathcal{S}$ runs over all qubit pairs.

The second part of the cost function, $C_2\left[\vec{\bm{\chi}}\right]$, encodes the CPTP constraints, and consists of the difference between the trace of each $\chi$-matrix estimate and the Hilbert space dimension (in this case 4), the sum of all negative eigenvalues of the $\chi$-matrix estimate (to constrain positivity), and the elements of Eq.~\eqref{eqn:CPcon}. The least-squares minimization solves the problem
\begin{align}
  \vec{\bm{\chi}}^{\rm est} = {\rm \arg}\min_{\vec{\bm{\chi}}}\left(C_1\left[\vec{\bm{\chi}}\right] + C_2\left[\vec{\bm{\chi}}\right]\right).
\end{align}
We note that this has the form of a semi-definite program (SDP). However, the operations involved in calculating $C_1$ are not obviously convex, and as a result the problem is not compatible with available convex-optimization packages. As such, we have not used this approach, but in future work hope to explore making the problem compatible with convex-optimization.

Even with the CPTP constraints imposed, the $\chi$-matrix estimates returned by the numerical solver will not necessarily be positive semi-definite. As such, we apply a post-processing step where we diagonalize each $\chi$-matrix estimate, generating a set of eigenvalues $\lambda_i$ with corresponding eigenvectors $\ket{v_i}$. We can then create a positive semi-definite $\chi$-matrix estimate for each two-qubit process
\begin{align}
  \tilde{\bm{\chi}}_{\mathcal{S}}^{\rm est} = \sum_{\lambda_i\geq0}\lambda_i\ketbra{v_i}{v_i}/\mathcal{N},
\end{align}
where $\mathcal{N}$ is a normilization factor to ensure ${\rm Tr}(\tilde{\bm{\chi}}_{\mathcal{S}}^{\rm est}) = 4$. These are what we use in the PAPA construction of the $N$-qubit gate.

Fig.~\ref{fig:CS} shows an example of the output from our implementation of the PAPA algorithm. For the CNOT gate with single-qubit decoherence described in the main text, it plots the difference between the measured (experimentally or in this case by simulation) Choi state, $\sigma_{12}$, and the PAPA reconstructed Choi state, $\rho_{12}$, for the effective process experienced by qubit pair 1-2. The element-wise difference is consistent with the magnitude of the trace distance reported in Fig.~\ref{fig:numsim} b).

Least-squares minimization requires an initial guess for the $\chi$-matrices, and we choose a decomposition of the ideal three-qubit gate as the initial guess. For the reconstructions presented in the main text, we found that their accuracy was mostly limited by numerical issues, such as the trade-off between the minimization tolerance and computation time. We observed a saturation in the trace distance for solver tolerance below a threshold value of $\epsilon_{\rm tol} = 10^{-7}$, which we attribute to the solver becoming stuck in a local minimum, see Fig.~\ref{fig:tol}.

In future work we hope to explore these numerical issues, and implement more efficient and accurate classical algorithms for the PAPA reconstruction. For instance, we would aim to prevent the solver from getting stuck in regions where the gradient of the cost function is below the tolerance threshold, but the solution accuracy is not. One route forward would be to adapt to PAPA more sophisticated optimization algorithms tailored for optimization over positive definite matrices, such as those using gradient descent \cite{Bolduc:2017aa,Knee:aa}.

\bibliography{PAPA_bib.bib}

\end{document}